# Phase Equilibria of Lattice Polymers from Histogram Reweighting Monte Carlo Simulations

Athanassios Z. Panagiotopoulos[*], Vicky Wong,

*School of Chemical Engineering,*
*Cornell University,*
*Ithaca, N.Y., 14850-5201, USA*

and M. Antonio Floriano

*Dipartimento di Chimica Fisica, Univ. Palermo*
*Via Archirafi 26, 90123 Palermo, ITALY*

[*] *To whom correspondence should be addressed. E-mail: azp2@cornell.edu*

**ABSTRACT**

Histogram-reweighting Monte Carlo simulations were used to obtain polymer / solvent phase diagrams for lattice homopolymers of chain lengths up to $r=1000$ monomers. The simulation technique was based on performing a series of grand canonical Monte Carlo calculations for a small number of state points and combining the results to obtain the phase behavior of a system over a range of temperatures and densities. Critical parameters were determined from mixed-field finite-size scaling concepts by matching the order parameter distribution near the critical point to the distribution for the three-dimensional Ising universality class. Calculations for the simple cubic lattice (coordination number $z=6$) and for a high coordination number version of the same lattice ($z=26$) were performed for chain lengths significantly longer than in previous simulation studies. The critical temperature was found to scale with chain length following the Flory-Huggins functional form. For the $z=6$ lattice, the extrapolated infinite chain length critical temperature is 3.70±0.01, in excellent agreement with previous calculations of the temperature at which the osmotic second virial coefficient is zero and the mean end-to-end distance proportional to the number of bonds. This confirms that the three alternative definitions of the Θ temperature are equivalent in the limit of long chains. The critical volume fraction scales with chain length with an exponent equal to 0.38±0.01, in agreement with experimental data but in disagreement with polymer solution theories. The width of the coexistence curve prefactor was tentatively found to scale with chain length with an exponent of 0.20±0.03 for $z = 6$ and 0.22±0.03 for $z = 26$. These values are near the lower range of values obtained from experimental data.

**INTRODUCTION**

Phase equilibria in polymer solutions are important in manufacturing, processing and applications of macromolecules. Significant progress has been made in recent years in the development of detailed atomistic models that can be used to predict properties of polymeric systems. Phase coexistence properties, however, are generally difficult to obtain directly for atomistically detailed models because the free energy of a system cannot be easily determined from simulations. Simple lattice models have often been used to obtain results that can be directly compared with statistical mechanical theories, but even for simple models, there is only a limited number of previous studies of coexistence properties for polymer/solvent systems.

Results for phase coexistence and critical properties of relatively short lattice homopolymers have been obtained previously by Yan et al.[1] on the $z=6$ simple cubic lattice for chains of length $r$ up to 200, by Mackie et al.[2] for the $z=26$ lattice for chains up to $r=128$ and by Wilding et al.[3] for the bond fluctuation model for chains up to $r=60$. Coexistence curves for continuous-space models were obtained by Sheng *et al.*[4] for a bead-spring model for chains up to $r=100$ and Escobedo and de Pablo[5] for square-well chains up to $r=100$. These previous studies have confirmed that the critical temperature, $T_c$, depends on chain length $r$ in a manner consistent with the functional form suggested by Flory-Huggins theory,

$$\frac{1}{T_c(r)} - \frac{1}{T_c(\infty)} \propto \frac{1}{\sqrt{r}} + \frac{1}{2r}, \tag{1}$$

where $T_c(\infty)$ is the critical temperature for chains of infinite length. For long chain lengths, equation 1 reduces to $T_c(\infty) - T_c(r) \propto r^{-x_3}$ with $x_3 = 0.5$, following the notation of reference[6].

Another important scaling relationship is that for the chain length dependence of the critical volume fraction, $\phi_c$,

$$\phi_c(r) \propto r^{-x_2}, \tag{2}$$

In previous lattice-based studies,[1-3] the exponent $x_2$ was found to be near or below the experimentally determined values[7,8] of 0.38 to 0.40 and significantly lower than the value $x_2 = 1/2$ suggested by Flory-Huggins theory. For the previous study of the continuous-space bead-spring model[4] the longest chain lengths studies were probably not in the scaling regime and the apparent exponent was significantly lower.

Finally, the chain-length dependence of the width of the coexistence curve near the critical point is expected to be described by[6],

$$\phi_1(r,T) - \phi_2(r,T) \propto r^{-x_1}(1 - T/T_c(r))^\beta, \tag{3}$$

where $\phi_1$ and $\phi_2$ are the volume fractions in the two coexisting phases and $\beta=0.326$ is the universal critical point exponent appropriate for three-dimensional fluid systems. The exponent $x_1$ has been determined experimentally[6,7] to be between 0.23 and 0.34 but no simulation estimates are available to the best of our knowledge.

In the present study, the histogram reweighting grand canonical Monte Carlo simulation technique[9] combined with mixed-field finite scaling concepts[10] has been employed. This technique has been recently applied by Wilding *et al.*[3] to calculate polymer/solvent critical point parameters of the bond fluctuation model of chains of length up to $r=60$. Because of the higher flexibility of the bond fluctuation model, these chain lengthss are equivalent to significantly longer ones on the simple cubic lattice.

The first part of the present manuscript deals with methodological issues related to the application of histogram-reweighting grand canonical Monte Carlo simulations to the prediction of phase diagrams and critical points of lattice homopolymers. The following section presents results for cubic lattices of coordination number $z=6$ and $z=26$. Results for the *z*=6 lattice are only in modest agreement with previous calculations[1] for chain lengths *r*=100 and 200. The infinite-chain length critical temperature is compared to independent estimates of the temperature at which chain dimensions behave ideally and the chain-chain second virial coefficient vanishes. We also obtain estimates of the exponents for scaling with chain length of the critical temperature, critical volume fraction and coexistence curve width and compare the results to available experimental data.

## SIMULATION METHODS

### HISTOGRAM-REWEIGHTING MONTE CARLO

The method has been described previously[9]; here, we would like to summarize the technique as applied to systems of interest to the present study. A grand canonical Monte Carlo simulation is performed in a simulation cell of size *V* under periodic boundary conditions, at an imposed value of the chemical potential $\mu$ and a temperature *T*. Particles are created and annihilated using the standard acceptance criteria.[11] The frequency of occurrence, *f(N,E)*, of *N* particles with total configurational energy *E* in the simulation cell is

$$f(N,E) = \Omega(N,V,E) e^{\beta(\mu N - E)} / \Xi(\mu,V,T), \quad (4)$$

where $\Omega(N,V,E)$ is the microcanonical partition function (density of states), $\beta$ is the inverse temperature ($\beta = 1/k_B T$, where $k_B$ is Boltzmann's constant), and $\Xi$ is the grand canonical partition function. Given the distribution function *f(N,E)*, collected in histogram form in the production period of a simulation, an estimate of the ratio of microcanonical partition functions for the system under study for two different values of *N* in the range covered by the simulation can be obtained directly as

$$\frac{\Omega_1(N_1,V,E_1)}{\Omega_2(N_2,V,E_2)} = \frac{f(N_1,E_1)}{f(N_2,E_2)} e^{\beta[\mu(N_1 - N_2) - (E_1 - E_2)]}. \quad (5)$$

In addition, one expects that a simulation at a different value of the chemical potential, $\mu'$, and temperature, $T'$, would result in a new distribution function, $f'(N,E)$, with

$$\frac{f'(N,E)}{f(N,E)} \propto e^{(\beta'\mu' - \beta\mu)N - (\beta' - \beta)E}. \quad (6)$$

The rescaling suggested by equation 6 can only be performed over a limited range of chemical potentials and temperatures since the original simulation provides statistically significant results only over a finite range of particle numbers and energies. For extending the range of particle numbers over which the partition function ratio can be determined from equation 5, several runs will need to be performed at different values of the chemical potential that result in overlapping distribution functions *f(N,E)*. From equation 4, the microcanonical partition function over the range of densities covered in each individual run, with index *n*, can be obtained from

$$\ln \Omega_n(N,V,E) = \ln[f(N,E)] - \beta\mu N + \beta E + C_n \quad (7)$$

where $C_n$ is a run-specific constant equal to the logarithm of the grand partition function for the chemical potential and temperature of run *n*, $\ln \Xi(\mu_n, V, T_n)$.

To obtain an estimate of the microcanonical partition function valid over a broad range of particle numbers and energies, results from different simulations need to be combined by assigning values of $C_n$ for each run in a self-consistent fashion. For combining results from multiple runs, the technique of Ferrenberg and Swendsen[12] is used. The probability $P(N,E;\mu,\beta)$ of a certain number of particles and a certain energy resulting by combining runs *n*=1 through R, assuming that they all have the same statistical efficiency is[12]

$$P(N,E;\mu,\beta) = \frac{\sum_{n=1}^{R} f_n(N,E) \exp[-\beta E + \beta\mu N]}{\sum_{m=1}^{R} K_m \exp[-\beta_m E + \beta_m \mu_m N - C_m]} \quad (8)$$

where $K_m$ is the total number of observations for run *m*. The constants $C_n$ are obtained from an iterative relationship:

$$\exp[C_n] = \sum_E \sum_N P(N,E;\mu_n,\beta_n) \quad (9)$$

The Ferrenberg-Swedsen method ensures that there is minimum deviation between observed and predicted histograms from the combined runs.

MIXED-FIELD FINITE-SIZE SCALING

In order to obtain critical parameter estimates, mixed-field finite size scaling methods[3] were used. A series of grand canonical simulations were performed near the expected critical point. The resulting histograms were combined according to equations 8 and 9 to obtain self-consistent estimates of the distribution functions $P(N, E; \mu, \beta)$. According to finite-size scaling theory, one needs to define an ordering operator, $\mathcal{M}$, combining the number of particles $N$ and energy $E$,

$$\mathcal{M} = N - sE \qquad (10)$$

where $s$ is a non-universal "field mixing" parameter controlling the strength of coupling between energy and density fluctuations near the critical point. At the critical point, the normalized probability distribution at a given system size $L = V^{1/3}$, $P_L(x)$, assumes a universal shape, with $x = a(L,r) \times (\mathcal{M} - \mathcal{M}_c)$. The non-universal scale factor $a(L,r)$ is chosen to result in unit variance for the distribution $P_L(x)$. An example of the matching of some of our data to the universal curve obtained from[13] is shown in Figure 1. There is excellent agreement between our data and the universal curve even though there are relatively few (<100) chains in the system, as also observed previously[3].

CONFIGURATIONAL-BIAS SAMPLING

For long polymers at moderate and higher densities, it is not practical to perform insertions and removals directly, as the probability of inserting a molecule in a random internal configuration without overlapping with existing particles decays exponentially with chain length. We have used the simple "athermal" version of configurational-bias sampling methods[14,15] to enable insertions and removals of the chain molecules. According to this scheme, the first monomer of a chain to be inserted is placed in a random position; if the position is occupied the attempted insertion fails. Subsequent monomers are placed on unoccupied positions on the lattice, provided such positions exist along the directions of growth permitted by lattice connectivity. The "Rosenbluth weight" for each growth step is calculated as the ratio of the number of unoccupied sites divided by the total number of sites, $z$=6 or 26 depending on the lattice. The Rosenbluth weight of the chain when it has been fully grown, $W_{new}$, is the product of the weights calculated during each growth step. The attempted insertion is accepted with probability that takes into account the energy change for the insertion and the Rosenbluth weight[11]. The reverse occurs during removal: a chain selected at random for a trial removal is "unzipped" from one randomly selected end to the other, and the Rosenbluth weight of the existing configuration is calculated as segment by segment is eliminated. We have chosen to use this limited form of configurational-bias sampling because it is computationally quite efficient.

MODELS AND SIMULATION DETAILS

In the model employed, homopolymers of chain length $r$ exist on a cubic lattice of coordination number $z$=6 or $z$=26. Monomeric solvent particles fill all space not occupied by polymer segments. For the $z$=6 case, non-bonded monomers interact when they are within one lattice spacing along the principal directions of the lattice: relative position vectors for interactions are (1,0,0) and the five additional vectors that result from symmetry operations in the -x, y, -y, z and -z directions. For the $z$=26 lattice, the relative position vectors for interactions are (1,0,0) (1,1,0), (1,1,1) and additional vectors that result from symmetry operations. In both cases, there is only a single relevant energy scale. If we denote by $\varepsilon_{PP}$, $\varepsilon_{PS}$ and $\varepsilon_{SS}$ the interaction energies for polymer-polymer, polymer-solvent and solvent-solvent interactions, the relevant energy scale is

$$\varepsilon = 2\varepsilon_{ps} - \varepsilon_{ss} - \varepsilon_{pp} \qquad (11)$$

The polymer-polymer interaction was set to -1 (resulting in attractive interactions for nearest-neighbor contacts) and the solvent-solvent and polymer-solvent interactions to zero. Temperature is normalized by the energy scale, so that $T^* = kT/\varepsilon$, where $k$ is Boltzmann's constant.

For all chain lengths studied, we made certain that the box length, *L*, was at least 4 times the maximum radius of gyration of the polymers for the temperatures and densities studied. We performed a mix of 50% particle creation/annihilation steps and 50% reptation steps.

Typically, after some preliminary runs to establish the approximate location of the critical point, we obtained histograms for a single long run at conditions near the estimated critical point. This was followed by additional, shorter runs near the expected liquid coexistence density at subcritical temperatures. The number of Monte Carlo steps per run was typically between 20 and 100 $\times 10^6$. Execution time required for a given number of steps is an increasing function of chain length, with a typical value being 1 hour / $10^6$ steps on a Intel Pentium 100 MHz processor. The acceptance ratio for the particle creation / annihilation step for runs near the critical point ranged from 65 % for *r*=32 on the *z*=26 lattice with mean volume fraction $<\varphi> = 0.20$, to 26 % for *r*=1000 on the *z*=6 lattice with mean volume fraction $<\varphi> = 0.066$. These acceptance ratios are significantly higher than those reported by Wilding *et al.* [3] for the bond fluctuation model. The difference is probably due to the simpler character of the cubic lattices we use and the less severe volume exclusion constraints of nearby monomers.

Data from histograms for a given lattice coordination number, chain length and system size were combined using the Ferrenberg-Swendsen algorithm[12] and analyzed to determine the critical point by matching to the universal order parameter distribution (see figure 1). Coexistence densities away from the critical point, for which the liquid and vapor peaks of the density distributions are clearly separated, were obtained by determining the value of the chemical potential at a certain temperature that results in equal areas of the vapor and liquid peaks. The coexistence densities were determined by the first moment of the corresponding peak. Near the critical point, where the liquid and vapor peaks of the density distributions overlap, coexistence densities were established by fitting to the scaling relationships

$$\frac{\phi_1 + \phi_2}{2} - \phi_c = A\tau^\mu \tag{12}$$

and

$$\phi_1 - \phi_2 = \tau^\beta (B + B_1 \tau^\theta) \tag{13}$$

where

$$\tau = \frac{T_c - T}{T_c} \tag{14}$$

The scaling exponents corresponding to the Ising universality class, $\beta=0.326$, $\mu=0.90$, and $\theta=0.54$, were employed. The constants $A$, $B$, and $B_1$ were identified as the parameters that yielded the best fit curves. The reduced distance from the critical point over which coexistence data were fitted to the scaling relationships was $\tau\approx 0.002$-$0.005$ for the longest chain lengths and $\tau\approx 0.01$-$0.05$ for the shortest chain lengths. Complete data sets are available on the world wide web[16].

Statistical uncertainties for selected systems and conditions were obtained by performing several duplicate runs and obtaining independent estimates for the critical parameters and coexistence densities.

RESULTS AND DISCUSSION

The first question that we needed to address is that of validation of our programs and analysis procedures. Some results for relatively short chains are available for the *z*=6 and *z*=26 lattices[1,2,17]. Our results are generally in agreement with previous literature data for the coexistence densities, within the stated uncertainties of the earlier results. However, some systematic discrepancies exist for the longer chain lengths between our data and those of Yan *et al*,[1] as shown in figure 2 for *r*=100. The coexistence liquid densities near the critical point are higher than the earlier results, but are probably within the statistical uncertainty of the latter, estimated as ±0.04 in liquid volume fraction by interpo-lating the error bars of figures 6 and 7 of[1]. Our results have statistical uncertainties smaller than the size of the symbols used in figure 2. The data of Madden *et al*[17] for the same system are in much better agreement with our calculations.

Coexistence data for chain lengths $r$=16 to $r$=1000 are shown in figures 3 and 4. Typical uncertainties of the calculated coexistence densities are ±1% of the reported liquid and vapor density. The estimated critical points as a function of chain length are shown on the figures by filled circles and reported in table 1. The reference point for the chemical potential reported in table 1 is the reversing random walk with no interactions. It should be emphasized that the critical point parameters are obtained for the specific simulation system size, $L$, reported in table 1. Typical statistical uncertainties for the critical parameters reported in table 1 are ±0.001 ($z$=6) and ±0.01 ($z$=26) for the temperature, ±0.001 for the critical volume fraction. The critical chemical potential is determined to within 4 significant figures for a fixed critical temperature. There is a high covariance of the critical chemical potential with the critical temperature, while the critical volume fraction is relatively insensitive to uncertainties in the critical temperature.

The dependence of the calculated critical parameters for two specific systems on simulation system size are shown in table 2. It is clear that there is a some residual dependence of the critical chemical potential on system size. The dependence of the critical temperature and volume fraction on system size is below the statistical uncertainty of the calculations. According to finite-size scaling theory[10] the critical temperature as a function of system size scales as

$$T_c^*(\infty) - T_c^*(L) \propto L^{-(\theta+1)/\nu} \qquad (15)$$

and the critical volume fraction scales approximately as

$$\langle \phi \rangle_c (L) - \langle \phi \rangle_c (\infty) \sim L^{-(1-\alpha)/\nu} \qquad (16)$$

Due to the presence of field-mixing in real fluids, only averages for the infinite-volume critical volume fraction can be obtained, as compared to direct estimate for the infinite-volume critical temperature. The scaling exponents in equations 15 and 16 are $\theta$=0.54, $\alpha$=0.11, and $\nu$=0.629.

The statistical uncertainties of our data are too great to allow for confirmation of the finite-size scaling functional dependence. For the purposes of analysis of the data and comparisons with theoretical and experimental estimates, it is sufficient to restrict our attention to data for a fixed system size, as reported in table 1.

Calculated critical temperatures are plotted as a function of chain length in figure 5. We have plotted $1/T_c$ *versus* $\frac{1}{\sqrt{r}} + \frac{1}{2r}$, as suggested by the Shultz-Flory relationship (equation 1). Linear regression of the data for $r \geq 64$ to obtain an estimate for the infinite chain length critical temperature yields $T_c(\infty)$=3.71±0.01 for $z$=6 and $T_c(\infty)$=20.85±0.01 for $z$=26. Linearity is followed very closely, with correlation coefficients $R^2 > 0.9999$ for both coordination numbers. The infinite chain length critical temperature estimate is not sensitive to the "cutoff" of lowest chain length included in the regression. For $z$=6, Yan *et al.*[1] report $T_c(\infty)$=3.45, a value 7% lower than the present estimate. In light of the agreement between our estimate for $T_c(\infty)$ and independent estimates of the Θ temperature discussed in the following paragraph, we conclude that the infinite-chain length critical temperature was previously[1] underestimated because of inaccurate critical temperatures for the longest chain lengths studied. For $z$=26, the present study is in reasonable agreement with the value $T_c(\infty)$=20.4 obtained by Mackie *et al*.

For the cubic lattice of coordination number $z$=6, Bruns[18] found that two definitions of the Θ temperature ($T_\Theta$) are equivalent in the limit of long chains, namely (a) the second osmotic virial coefficient is equal to zero and (b) the mean square end-to-end distance is proportional to the number of bonds. The common value was obtained[18] as $T_\Theta$ = 3.713. Our estimate of the critical temperature in the limit of infinite chain length ($T_c(\infty)$=3.71±0.01) coincides with this value, thus confirming a long-standing premise that *all three* definitions of the Θ temperature are equivalent in the limit of long chain lengths. This is the first time that this important assumption of polymer solution theories is confirmed at a level of less than 0.3% uncertainty. A previous calculation[4] for a continuous space model obtained agreement between the three definitions of the Θ temperature to within 4%.

Scaling of critical volume fraction with chain length is shown in figure 6. For $z=6$, regression of the data for $r \geq 64$ gives a slope of $x_2=0.36\pm0.02$, and for $z=26$, $x_2=0.39\pm0.02$. These values are in good agreement with experimental measurements[7,8] that give a range for the exponent $x_2$ between 0.38 and 0.40. These values are clearly significantly lower than the Flory-Huggins prediction of $x_2=0.50$. The only previous simulation study[3] to obtain a value for the exponent $x_2$ comparable to the experimental value was by Wilding *et al* for the bond fluctuation model, yielding $x_2=0.369$. Previous studies[1,2] for the $z=6$ and $z=26$ simple cubic lattices obtained significantly lower values for the exponent $x_2$, primarily because they were restricted to shorter chain lengths for which the effective exponent is lower, as can be seen in figure 6.

Finally, a quantity of significant interest in polymer solution theories is the exponent $x_1$ for scaling of the width of the coexistence curve with chain length according to equation 3. Figure **7** shows the results of our calculations for the quantity $B$ defined *via* equation 13, as a function of chain length. The slope of the lines in figure **7** are equal to the exponent $x_1$. There is significant scatter in the data of figure **7** and the exponent value depends on the range of data chosen for the regression. A possible explanation for the relatively poor data quality is that, due to the limitations on system size that we have used, we can only obtain phase coexistence information up to a sizable reduced distance from the critical point of $\tau \geq 0.001$ (for long chains) to $\tau \geq 0.01$ (for short chains). This may be too far for reliable extrapolations from the scaling relationships (equations 12-14). Despite this limitation, we felt it was worth analyzing the data to obtain a value for the exponent $x_1$. We have chosen to use the data for chain lengths $r \geq 100$ for the $z=6$ lattice, which yield $x_1=0.20\pm0.03$, and the data for $r \geq 64$ for the $z=6$ lattice, which give $x_1=0.22\pm0.03$. The experimental value quoted by Dobashi *et al*.[7] is $x_1=0.23$. Sanchez[19] reanalyzed the data of Dobashi *et al.* and quoted 0.28 as the value of the exponent. Shinozaki *et al.*[8] find $x_1=0.34$ from analysis of their own data. From theoretical considerations,[20] the exponent $x_1$ takes on the value 0.25 in mean-field theory, and a value of $(1-\beta)/2=0.34$ according to de Gennes' scaling argument.[21] It is clear that additional simulation work will be required to obtain an accurate value of the exponent $x_1$. Such work will probably need to utilize larger system sizes to permit direct calculation of coexistence data in the vicinity of the critical point and should also investigate longer chain lengths to ensure that the long-chain limit has been reached.

CONCLUSIONS

We have used grand canonical Monte Carlo simulations combined with histogram reweighting techniques to obtain phase coexistence properties in polymer/solvent systems. Simple cubic lattices of coordination number $z=6$ and $z=26$ were investigated for chain lengths up to $r=1000$. This range of chain lengths significantly exceeds the range of previous simulation studies for comparable systems.

Critical temperatures were found to scale with chain length in accordance to the Shultz-Flory prediction. This is in agreement with previous simulation studies of continuous-space and lattice model homopolymers. For the $z=6$ lattice, the extrapolated infinite chain length critical temperature is $3.71\pm0.01$, in excellent agreement with previous calculations of the temperature at which the osmotic second virial coefficient is zero and the mean end-to-end distance proportional to the number of bonds. This confirms, to an unprecedented level of accuracy, the standard assumption of polymer theories that all three definitions of the Θ temperature are equivalent at the limit of long chains.

Critical volume fractions were found to scale with chain length with exponent $x_2=0.38$, in excellent agreement with experimental data, but in disagreement with most polymer solution theories. Almost all previous simulations of polymer-solvent coexistence curves yielded lower values for this exponent because the chain lengths investigated were not in the scaling regime.

We were unable to reach definite conclusions about the scaling with chain length of the prefactor to the width of the coexistence curve, because of statistical uncertainties of our data. These uncertainties are likely the result of insufficient approach to the critical point due to limitations in system size. From our data, we obtain a value of the exponent $x_1$ near the lower range of the experimentally observed range $x_1 \approx 0.23$-$0.34$. Additional simulation work with larger system sizes is needed to clarify the situation with respect to this exponent.

The simulation methods we have used can be used to study polymeric systems of even longer chain lengths at the vicinity of critical points. The reason for this is that the critical volume fraction is lower for longer chains, thus allowing reasonable statistics for the grand canonical insertions and removals. The methods are also applicable to mixtures of different polymers or solvents.

ACKNOWLEDGMENTS

Research on which this work was based was supported by a grant from U.S. Department of Energy, Office of Basic Energy Sciences. M.A.F. would like to acknowledge travel support by a NATO Senior Research Fellowship. We would like to thank Dr. Nigel Wilding for helpful discussions, providing preprints of papers prior to publication and providing the data for the universal Ising distribution and Prof. Ben Widom for helpful discussions and comments on the manuscript prior to submission.

14

15

Table 1. Critical parameters as a function of chain length and lattice coordination number. See text for estimates of the statistical uncertainties of the results.

| | z=6 | | | | z=26 | | | |
|---|---|---|---|---|---|---|---|---|
| r | L | $T_c(L)$ | $\phi_c(L)$ | $\mu_c(L)$ | L | $T_c(L)$ | $\phi_c(L)$ | $\mu_c(L)$ |
| 8 | 15 | 2.147 | 0.359 | -9.182 | 15 | 11.87 | 0.306 | -60.45 |
| 16 | 17 | 2.468 | 0.296 | -10.01 | 20 | 13.74 | 0.247 | -82.63 |
| 32 | 20 | 2.749 | 0.248 | -7.119 | 20 | 15.36 | 0.201 | -104.17 |
| 64 | 32 | 2.982 | 0.199 | 4.340 | 30 | 16.71 | 0.160 | -119.81 |
| 100 | 30 | 3.106 | 0.173 | 20.49 | 40 | 17.42 | 0.136 | -122.42 |
| 200 | 40 | 3.266 | 0.137 | 71.79 | 50 | 18.33 | 0.103 | -102.21 |
| 400 | 50 | 3.387 | 0.107 | 183.7 | 60 | 19.03 | 0.080 | -23.24 |
| 600 | 65 | 3.443 | 0.093 | 300.3 | 70 | 19.35 | 0.068 | 73.57 |
| 800 | 75 | 3.477 | 0.080 | 418.9 | 75 | 19.54 | 0.061 | 177.9 |
| 1000 | 85 | 3.503 | 0.072 | 539.8 | 85 | 19.68 | 0.056 | 287.5 |

Table 2. Critical parameters for selected systems as a function of simulation system size.

| L | $T_c(L)$ | $\phi_c(L)$ | $\mu_c(L)$ |
|---|---|---|---|
| | z=26, r=600 | | |
| 55 | 19.34 | 0.068 | 73.00 |
| 65 | 19.34 | 0.069 | 73.23 |
| 70 | 19.35 | 0.068 | 73.57 |
| 95 | 19.35 | 0.067 | 73.78 |
| | z=26, r=800 | | |
| 75 | 19.54 | 0.061 | 177.9 |
| 87 | 19.54 | 0.062 | 178.2 |
| 110 | 19.54 | 0.060 | 178.3 |

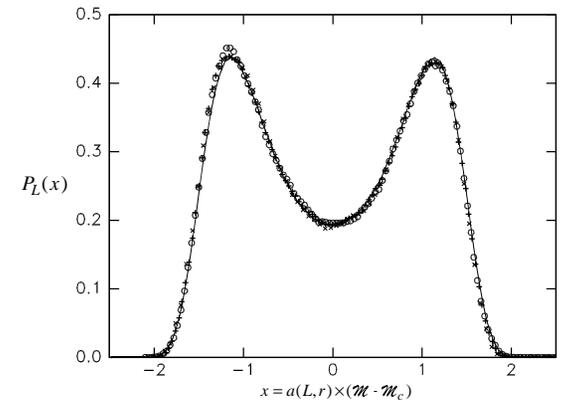

Fig. 1. Matching of the scaled order parameter distribution to the universal curve for the Ising three-dimensional universality class, indicated by the continuous line. Points are from our simulations for z=26: (+) r=64, L=30; (×) r=200, L=50; (○) r=600, L=95.

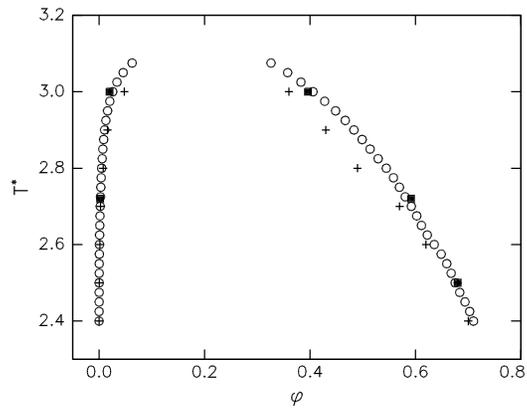

Fig. 2. Phase diagram for *z*=6, *r*=100. (○) This work; (■) Madden *et al.*[17]

(+) Yan *et al.*[1]

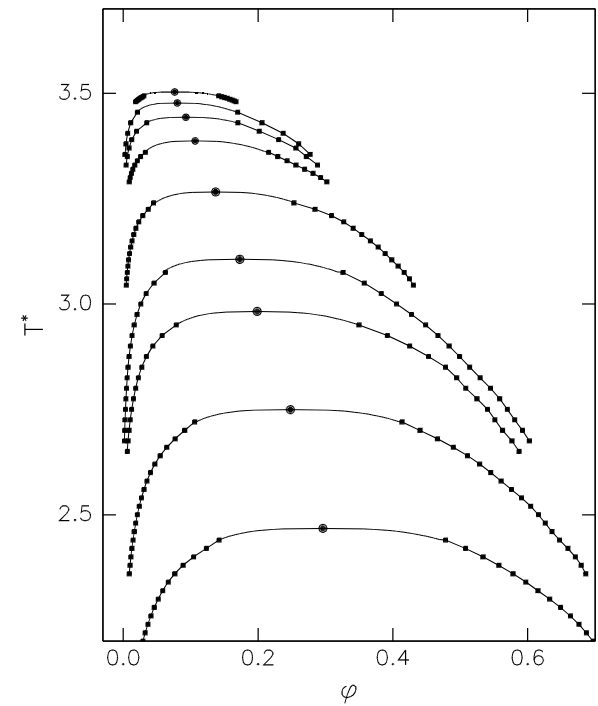

Fig. 3. Calculated phase diagrams for *z*=6. The estimated location of the critical points is given by (●) and the directly measured coexistence data by (■). Lines connect the measured points and are extrapolated to the critical point using equations 12-14. From top to bottom, the curves correspond to *r*=1000, 800, 600, 400, 200, 100, 64, 32, 16.

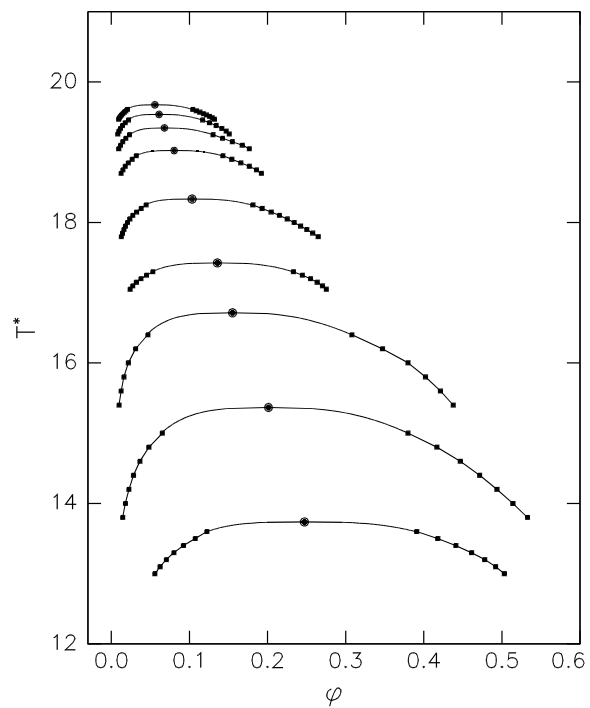

Fig. 4. Calculated phase diagrams for $z$=26. The estimated location of the critical points is given by (●) and the directly measured coexistence data by (■). Lines connect the measured points and are extrapolated to the critical point using equations 12-14. From top to bottom, the curves correspond to $r$=1000, 800, 600, 400, 200, 100, 64, 32, 16.

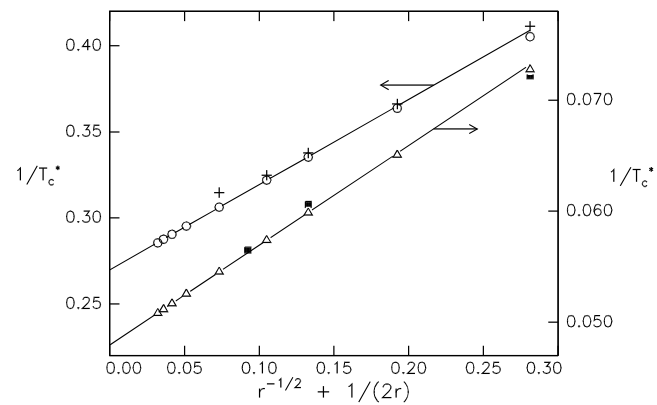

Fig. 5. Scaling of critical temperature with chain length. Left axis, $z$=6: (○) This work; (+) Yan et al.[1]; Right axis, $z$=26: (△) This work; (■) Mackie et al.[2] Lines are fitted to the critical temperatures from this work, $r \geq 64$.

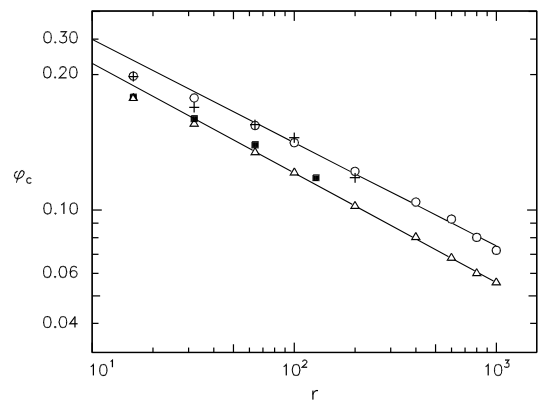

Fig. 6.  Scaling of critical volume fraction with chain length.  Top, $z=6$: (○) This work; (+) Yan *et al.*[1]; Bottom, $z=26$: (△) This work; (■) Mackie *et al.*[2]  Lines are fitted to the critical volume fractions from this work, $r \geq 64$.

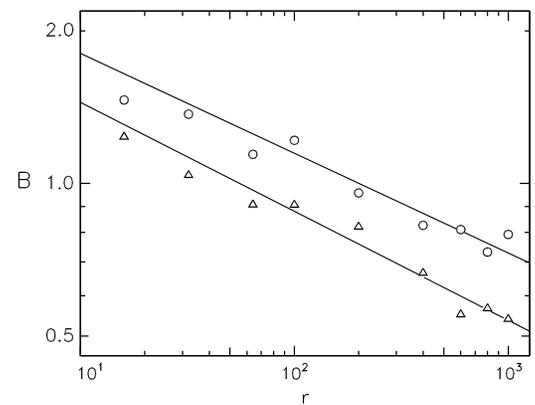

Fig. 7.  Scaling of parameter B with chain length from the present work.  Top line, (○) $z=6$; Bottom line, (△) $z=26$.   Lines are fitted to the coexistence curves from this work, $r \geq 100$ ($z=6$) and $r \geq 64$ ($z=26$).